\documentclass[12pt]{amsart}
\usepackage{amsthm}
\usepackage[
  top=1.15in,
  bottom=1.15in,
  left=1.25in,
  right=1.25in,
  heightrounded,
]{geometry}
\usepackage{txfonts}

\newtheorem{theorem}{Theorem}
\newtheorem{definition}{Definition}

\begin{document}

\title[The Nonlocal $\overline{\partial}$-Problem and (2+1)D Integrable Systems]{On Solutions to the Nonlocal $\overline{\partial}$-Problem and (2+1) Dimensional Completely Integrable Systems}
\author{Patrik V. Nabelek}
\date{\today}
\maketitle
\maketitle
\begin{abstract} In this short note we discuss a new formula for solving the nonlocal $\overline{\partial}$-problem, and discuss application to the Manakov--Zakharov dressing method. We then explicitly apply this formula to solving the complex (2+1)D  Kadomtsev--Petviashvili equation and complex (2+1)D completely integrable generalization of the (2+1)D Kaup--Broer (or Kaup--Boussinesq) system. We will also discuss how real (1+1)D solutions are expressed using this formalism. It is simple to express the formalism for finite gap primitive solutions from \cite{NZ20}, \cite{N20} using the formalism of this note. We also discuss recent results on the infinite soliton limit for the (1+1)D Korteweg--de Vries equation and the (2+1)D Kaup--Broer system. In an appendix, the classical solutions to the 3D Laplace equation (2+1)D d'Alembert wave equation by Whittaker are described. This appendix is included to elucidate an analogy between the dressing method and the Whittaker solutions. 
%\keywords{Completely Integrable System \and The Nonlocal $\overline{\partial}$-Problem \and The KdV Equation \and The KP Equation \and The Kaup--Broer System}
\end{abstract}

\section{Introduction}

The KdV equation is
\begin{equation} 4u_t -6 u u_x + u_{xxx} = 0. \end{equation}
The KdV equation described weakly nonlinear long wave propagating in (1+1)D, and is the first known example of a completely integrable nonlinear PDE.
The KP equation
\begin{equation} (4u_t - 6 u u_x + u_{xxx})_x + \alpha^2 u_{yy} = 0\end{equation}
is a two dimensional completely integrable generalization of the KdV equation, and describes weakly nonlinear long waves in (2+1)D that are primarily traveling in the $x$ direction, with small deviations in the $y$ dimension. Solutions to the KP equation that do not satisfy this condition are not physically realized in weakly nonlinear long wave systems.
From the point of view of a real PDE, there are two scaling classes corresponding to $\alpha=1$ and $\alpha=i$ called the KP-2 and KP-1 equation respectively. However, if we allow the independent variables to take complex values there is a single scaling class $\alpha=1$ since we can evaluate the solutions on the line $y = i \upsilon$ with $\upsilon \in \mathbb{R}$. 

The KP equation can be solved by the inverse scattering problem using a nonlocal $\overline{\partial}$-problem \cite{ABF83}. The structure of the nonlocal $\overline{\partial}$-problem was further used by Zakharov and Manakov to discuss inverse scattering for general classes of problems in dimensions higher than (1+1)D \cite{ZM84,ZM85}. This formulation of the inverse scattering problem also leads to a class of solutions to the KdV equation called primitive solutions that have been studied in \cite{DZZ16,DNZZ20,GGJM18,N20,NZ20,NZZ18,ZDZ16,ZZD16}.

The four scaling classes of the (1+1)D Kaup--Broer system can be represented as
\begin{align}
&\varphi_t + \frac{1}{2}  (\varphi_x)^2 + \varepsilon \eta = 0, \\
&\eta_t + \varphi_{xx} + (\eta \varphi_x)_x + \mu \varphi_{xxxx} = 0,
\end{align}
were $\varepsilon = \pm 1$ and $\mu = \pm 1/4$.
The (1+1)D Kaup--Broer equation was introduced by Kaup and Broer as a completely nonlinear system that described shallow water long waves.
The Cauchy problem for the (1+1)D Kaup--Broer system is ill-posed because of instabilities that lead to finite time blow up \cite{ABF83}.
However, a class of solutions called primitive solutions introduced in \cite{NZ20} see, to be stable and bounded and thus could potentially still be analyzed in practical models even if the full space of solutions is problematic.

A complex (2+1)D completely integrable generalization of the KB system is
\begin{align}
 &\varphi_t + \alpha \varphi_u^2-\beta \varphi_{v}^2 + \Pi = 0 \\
& \eta_t+2\alpha (\varphi_u \eta)_u - 2\beta(\varphi_v \eta )_v  - 2 a \alpha \varphi_{uu} + 2 a \beta \varphi_{vv} + \frac{\alpha}{2}  \varphi_{uuuv} - \frac{\beta}{2} \varphi_{uvvv} = 0 \\
 & \Pi_{uv} = 2 \alpha \eta_{uu} - 2 \beta \eta_{vv}.
\end{align}
A real version of the system was introduced in \cite{RP11}.
The nonlocal $\overline{\partial}$-dressing method for solving the fully complex version of this equation was discussed in \cite{NZ20}.
The real version introduced by \cite{RP11} generalizes a single canonical scaling of the Kaup--Broer system, while the fully complex version generalizes all four scaling classes of the Kaup--Broer system \cite{NZ20}.

The KdV equation is an isospectral evolution of 1D Schr\"{o}dinger operators.
The KP equation is an isospectral evolution of the (1+1)D Schr\"{o}dinger equation with a fixed spectral curve.
The (2+1)D generalization of the Kaup--Broer system is an isospectral evolution of 2D Schr\"{o}dinger equations at fixed energy level---for finite gap solutions, this would mean it preserves the spectral curve \cite{NZ20}.

These equations can be solved using the dressing method.
The dressing method is a name given to a whole family of methods for solving completely integrable partial differential equations.
The formulation of the dressing method addressed in this paper is based on the nonlocal $\overline{\partial}$-problem method, and we summarize this as our main theorem.

However, before we state the theorem, we need to introduce some distribution spaces of distributions.
The theorem and formal proof extend to broader classes of distributions, but the following is sufficient for the purposes of this note.
\begin{definition}
We introduce a distribution $R(\lambda,\eta)$ called the dressing function of the form
\begin{align} R(\zeta,\eta) = & \sum_{n = 1}^{N_0} r_{n} \delta(\zeta-\zeta_{n}) \delta(\eta-\eta_{n}) \\ & + \sum_{m = 1}^{N_1} \int_{\gamma_{m}} \tilde R_{m}(s) \delta(\lambda-\tilde \zeta_{m}(s)) \delta(\eta - \tilde \eta_{m}(s)) ds \\
& + \sum_{\ell = 1}^{N_2} \iint_{\Sigma_{\ell}} \breve R_{\ell}(p) \delta(\zeta - \breve \zeta_{\ell}(p))  \delta(\eta - \breve \eta_{\ell}(p)) d^2 p \end{align}
where: $\zeta_n, \eta_n$ are complex numbers; $\gamma_m$ are intervals, $\tilde R_m$ are H\"{o}lder continuous  and $\tilde \zeta_m, \tilde \eta_m$ are embedding of $\gamma_m$ into $\mathbb{C}$; $\Sigma_n$ are two dimensional submainfolds of $\mathbb{R}^2$, $\breve R_n$ are H\"{o}lder continuous functions on $\Sigma_n$, and $\breve \zeta_n$ and $\breve \eta_n$ are embeddings of $\Sigma_n$ into $\mathbb{C}$.
For a choice of dressing function $R$ of this form, we define the space $\mathcal{D}$ of distributions of the form
\begin{align} f(\zeta) = & \sum_{n=1}^{N_0} f_n \delta(\zeta-\zeta_n) + \sum_{m = 1}^{N_1} \int_{\gamma_m} \tilde f_m(s) \delta(\zeta - \tilde \zeta_m(s)) ds \\
& + \sum_{\ell = 1}^{N_2} \iint_{\Sigma_\ell} \breve f_\ell(p) \delta(\zeta-\breve \zeta_\ell(p)) d^2p  \end{align}
where $f_n$ are numbers, $\tilde f_n$ are H\"{o}lder continuous functions on $\gamma_n$, and $\breve f_m$ are H\"{o}lder continuous functions on $\Sigma_n$.
\end{definition}

\begin{theorem}
If $f$ solves the system
\begin{equation} f(\zeta) - \iint_{\mathbb{C}} K(\zeta,\xi) f(\xi)d^2 \xi = g(\zeta), \end{equation}
where
\begin{equation} K(\zeta,\xi) = \frac{1}{\pi} \iint_{\mathbb{C}} \frac{R(\zeta,\eta)}{\eta - \xi} d^2 \eta, \end{equation}
and
\begin{equation} g(\zeta) = \iint_{\mathbb{C}} R(\zeta,\eta) d^2 \eta, \end{equation}
assuming the operator $\mathcal{L}:\mathcal{D} \to \mathcal{D}$ defined by
\begin{equation} \mathcal{L} f (\zeta) = f(\zeta) - \iint_{\mathbb{C}} K(\zeta,\xi) f(\xi)d^2 \xi  \end{equation}
is welldefined with trivial nullspace,
then
\begin{equation} \chi(\lambda) = 1 + \frac{1}{\pi} \iint_{\mathbb{C}} \frac{f(\zeta)}{\lambda - \zeta} d^2 \zeta \end{equation}
is the unique solution to the nonlocal $\overline{\partial}$-problem
\begin{equation} \frac{\partial \chi}{\partial \bar \lambda} (\lambda) = \iint_{\mathbb{C}} R(\lambda,\eta) \chi(\eta) d^2 \eta. \end{equation}
satisfying $\chi(\lambda) = 1 + O(\lambda^{-1})$ as $\lambda \to \infty$.
\end{theorem}

Applying the main theorem to the nonlocal $\overline{\partial}$-problem dressing method for the KP equation gives the following form 
\begin{equation} u(x,y,t) = \frac{2}{\pi} \frac{\partial}{\partial x} \iint_{\mathbb{C}} f(\xi,x,y,t) d^2\xi \end{equation}
of a solution to the (complex) KP equation, where  $f(\lambda)$ is the solution to
\begin{equation}  f(\zeta) - \frac{1}{\pi} \iiiint_{\mathbb{C}^2} \frac{R(\zeta,\eta)}{\eta-\xi} f(\xi) d^2 \eta  d^2 \xi  =   \iint_{\mathbb{C}} R(\zeta,\eta) d^2 \eta,   \end{equation}
and
\begin{equation} R(\lambda,\eta) = e^{\phi(\eta,x,y,t) - \phi(\lambda,x,y,t)} R_0(\lambda,\eta) \end{equation}
where $\phi(\lambda,x,y,t) = \lambda x +  \lambda^2 y + \lambda^3 t$
and $R_0$ is independent of $x$, $y$, $t$ (there is an error here, see footnote \footnote{The definition of the function $\phi(\lambda,x,y,t)=\lambda x+\lambda^2y + \lambda^3 t$ on the lines following equations (21) and (49) lead to solutions to the scaling
$$(4u_t+6uu_x-u_{xxx})_x - 3u_{yy} =0$$
of the complex KP equation. 
The real version of this complex KP equation is still the KP-2 equation. To produce solutions to the scaling
$$(4u_t-6uu_x+u_{xxx})_x + u_{yy} = 0$$
used in this note, the correct definition to use is
$$\phi(\lambda,x,y,t) = \lambda x + \sqrt{3}\lambda^2 y - \lambda^3 t.$$
When this scaling is used $t_2=\sqrt{3}y$ and $t_3=-t$ in (52).

{\it This correction is fairly involved, so it is given as a footnote to maintain continuity between the preprint and the published version. A couple other typos were corrected within the text.}}).
We will also discuss the computations of isospectral manifolds containing all potentials with the same spectrum.
The solutions to completely integrable systems are dynamical systems on the isospectral manifolds.

\section{The Nonlocal $\overline{\partial}$-Problem}

A nonlocal $\overline{\partial}$-problem is an integro-differential equation of the form
\begin{equation} \frac{\partial \chi}{\partial \bar \lambda}(\lambda) = \iint_{\mathbb{C}} R(\lambda, \eta) \chi(\eta) d^2 \eta \end{equation}
where $R(\lambda,\eta)$ is a distribution of the form introduced in definition 1, and $d^2 \eta$ is the Lebesgue measure on $\mathbb{R}^2 \equiv \mathbb{C}$.
We look for a solution satisfying the boundary condition $\chi(\lambda) = 1+ O(\lambda^{-1})$ as $\lambda \to \infty$.
Suppose the support of $R(\lambda,\eta)$ is some embedded compact real sub-manifold of $\mathbb{C}^2$.

The solution $\chi(\lambda)$ to the nonlocal $\overline{\partial}$-problem solves the integral equation
\begin{equation} \chi(\lambda) = 1 + \frac{1}{\pi} \iiiint_{\mathbb{C}^2} \frac{R(\zeta, \eta) \chi(\eta)}{\lambda - \zeta} d^2 \zeta d^2 \eta .  \end{equation}
We look for a solution $\chi(\lambda)$ to this integral equation of the form
\begin{equation} \chi(\lambda) = 1 + \frac{1}{\pi} \iint_{\mathbb{C}} \frac{f(\xi)}{\lambda - \xi} d^2 \xi.\end{equation}
Then the generalized function $f(\lambda,\zeta)$ solves the singular integral equation
\begin{equation}  \iint_{\mathbb{C}} \frac{f(\xi)}{\lambda - \xi}  d^2 \xi - \frac{1}{\pi} \idotsint_{\mathbb{C}^3} \frac{R(\zeta,\eta) f(\xi)}{(\lambda-\zeta)(\eta - \xi)}  d^2 \eta  d^2 \xi d^2 \zeta =   \iiiint_{\mathbb{C}^2} \frac{R(\zeta,\eta)}{\lambda-\zeta} d^2 \eta d^2 \zeta   \end{equation}
where $R$ and $f$ are generalized functions supported on a compact subset of $\mathbb{C}^2$ that has regularity no worse than delta functions.
This is an integral equation satisfied by a charge density that is supported on a set of points in the plane, a set of contours in the plane, and a set of 2D subsets of the plane.

If the integral transform
\begin{equation} \mathcal{H}_{\mathbb{C}} f(\lambda) = \iint_{\mathbb{C}} \frac{f(\xi)}{\lambda - \xi} d^2 \xi \end{equation}
is invertible on an appropriate space of generalized functions for a specific application, then this singular integral equation takes the form
\begin{equation}   f(\lambda) -  \iint_{\mathbb{C}} K(\lambda,\xi)  f(\xi)  d^2 \xi  =   g(\lambda) \end{equation}
where the generalized function $K(\lambda,\xi)$ is the kernel
\begin{equation} K(\lambda,\xi) = \frac{1}{\pi} \iint_{\mathbb{C}} \frac{R(\lambda,\eta)}{\eta-\xi} d^2 \eta \end{equation}
and the generalized function $g(\lambda)$ is
\begin{equation} g(\lambda) = \iint_{\mathbb{C}} R(\lambda,\eta) d^2 \eta. \end{equation}
Each choice of $R$ leads to a choice of $K$ and $g$, so we only need to consider a case by case basis.

The solution to the nonlocal $\overline{\partial}$-problem
\begin{equation} \frac{\partial \chi}{\partial \bar \lambda} (\lambda) = \iint_{\mathbb{C}} R(\lambda,\eta) \chi(\eta) d^2 \eta \end{equation}
is then given by
\begin{equation} \chi(\lambda) = 1 + \frac{1}{\pi} \iint_{\mathbb{C}} \frac{f(\xi)}{\lambda-\xi} d^2 \xi. \end{equation}
In summary, the nonlocal $\overline{\partial}$-problem is equivalent to the singular integral equation
\begin{equation}   f(\lambda) -  \iint_{\mathbb{C}} K(\lambda,\xi)  f(\xi)  d^2 \xi  =   g(\lambda) \end{equation}
where the generalized function $K(\lambda,\xi)$ is the kernel
\begin{equation} K(\lambda,\xi) = \frac{1}{\pi} \iint_{\mathbb{C}} \frac{R(\lambda,\eta)}{\eta-\xi} d^2 \eta \end{equation}
and the generalized function $g(\lambda)$ is
\begin{equation} g(\lambda) = \iint_{\mathbb{C}} R(\lambda,\eta) d^2 \eta. \end{equation}

This form of the integral equation produces complex and singular solutions to the KP equation.
To produce a real solution, a reality condition needs to be introduced that will imply that $R(\lambda,\zeta)$ has to be supported on a surface.
Similarly, the dimensional reduction of a (2+1)D integrable system to (1+1)D will involve a similar condition that $R(\lambda,\zeta)$ has to be restricted to a surface.
The intersection of these surfaces will be a real curve, and this is why the dressing functions for solutions to real (1+1)D equations are supported on curves.

\subsection{Rational Solutions to the Nonlocal $\overline{\partial}$-Problem} To produce rational solutions we make the assumption that
\begin{equation} R(\zeta,\eta) = \sum_{n = 1}^N r_n \delta(\zeta-\zeta_n) \delta(\eta-\eta_n),\end{equation}
\begin{equation} f(\xi) = \sum_{n = 1}^N f_n \delta(\xi-\zeta_n),
\end{equation}
where $\delta$ is the Dirac delta function on the plane, and $\eta_n$.
Then $f_n$ solves
\begin{equation} f_n - \frac{1}{\pi} \sum_{m = 1}^N \frac{r_n}{\eta_n-\zeta_m} f_m = r_n.  \end{equation}
The values $f_n$ are determined from $\{r_n,\zeta_n,\eta_n\}$ by solving this linear equation.
A subset of these solutions lead to multi-soliton solutions to integrable systems, such as the multi-soliton solutions to the KdV equation and KP equation.
In practice, reality and regularity conditions on $r_n$, $\zeta_n$ and $\eta_n$ must be enforced to make sure multi-soliton solutions are produced.

As an example, the dressing functions of the form
\begin{equation} R(\zeta,\eta) = \sum_{n = 1}^N r_n e^{-2\kappa_n x + 2\kappa_n^3 t} \delta(\zeta-\kappa_n) \delta(\eta+\kappa_n)\end{equation}
with $r_n>0$ and $\kappa_n>0$ allow computation of all N-soliton solutions to the KdV equation.

\subsection{Sectionally Holomorphic Solutions to the Nonlocal $\overline{\partial}$-Problem}

Let us consider
\begin{equation} R(\lambda,\eta) = \sum_{n = 1}^N \int_{\gamma_n} \tilde R_n(s) \delta(\lambda-\tilde \zeta_n(s)) \delta(\eta - \tilde \eta_n(s)) ds  \end{equation}
where $\gamma_n$ are a collection of intervals, the functions $\tilde R_n$ are H\"{o}lder continuous functions on $\gamma_n$, and $\tilde \zeta_n$ and $\tilde \eta_n$ are embeddings on $\gamma_n$ into $\mathbb{C}$.
We consider $f$ of the form
\begin{equation} f(\zeta) = \sum_{n = 1}^N \int_{\gamma_n} \tilde f_n(s) \delta(\zeta - \tilde \zeta_n(s)) ds \end{equation}
where $\tilde f_n$ are H\"{o}lder continuous functions on $\gamma_n$.
The functions $\tilde f_n(s)$ solves the system of integral equations
\begin{equation} \tilde f_n(s) - \frac{1}{\pi} \sum_{m = 1}^N \int_{\gamma_m} \frac{ \tilde R_n(s) \tilde f_m(s')}{\tilde \eta_n(s')- \tilde \zeta_m(s)} ds' =\tilde R_n(s). \end{equation}
The functions $\tilde f_n$ are determined from $\{\tilde R_n,\tilde \zeta_n,\tilde \eta_n\}$ by solving this integral equation.

These solutions can also be computed using a nonlocal scalar Riemann--Hilbert problem, and an equivalent local vector Riemann--Hilbert problem.

This case includes all primitive solutions to the KdV equation and the KB system \cite{DNZZ20,DZZ16,GGJM18,N20,NZ20,NZZ18,ZDZ16,ZZD16}.
Moreover, if we allow one of the intervals $\gamma_m$ to be infinite, the generalized primitive solutions considered in \cite{ZZ19} can be computed. In particular, all finite gap solutions of the KdV equation and likely all physically relevant finite gap solutions to the KB system can be produced in this manner \cite{N20,NZ20}.

For example, the dressing functions of the form
\begin{align} R(\lambda,\eta) = & \sum_{n = 1}^{g} \int_{\kappa_{2n-1}}^{\kappa_{2n}} e^{-2sx+2s^3t+\sum_{j=1}^g a_j s^{2j-1}} \delta(\lambda-s) \delta(\eta +s)  ds \\
& + \sum_{n = 1}^{g} \int_{\kappa_{2n-1}}^{\kappa_{2n}} e^{2sx-2s^3t-\sum_{j=1}^g a_j s^{2j-1}} \delta(\lambda+s) \delta(\eta -s)  ds \end{align}
for fixed $a_j \in \mathbb{R}$ and $\kappa_j>0$ allows the production of all finite gap solutions to the KdV equation with spectral curve
\begin{equation} w^2 = \lambda(\lambda + \kappa_1^2)(\lambda+\kappa_2^2)\dots(\lambda + \kappa_{2g}^2). \end{equation}

\subsection{Solutions to the Nonlocal $\overline{\partial}$-Problem Supported on 2D Subsets of $\mathbb{C}$}

Let us consider
\begin{equation} R(\zeta,\eta) = \sum_{n = 1}^N \iint_{\Sigma_n} \breve R_n(p) \delta(\zeta - \breve \zeta_n(p))  \delta(\eta - \breve \eta_n(p)) d^2 p \end{equation}
where $\Sigma_n$ are two dimensional submainfolds of $\mathbb{R}^2$, $\breve R_n$ are H\"{o}lder continuous functions on $\Sigma_n$, and $\breve \zeta_n$ and $\breve \eta_n$ are embeddings of $\Sigma_n$ into $\mathbb{C}$. 
We consider $f$ of the form
\begin{equation} f(\zeta)= \sum_{n = 1}^N \iint_{\Sigma_n} \breve f_n(p) \delta(\zeta-\breve \zeta_n(p)) d^2p.  \end{equation}
The functions $\breve f_n(p)$ solve
\begin{equation} \breve f_n(p) - \frac{1}{\pi} \sum_{m =1}^N \iint_{\Sigma_n}\frac{\breve R_n(p) \breve f_m(p')}{\breve \eta_n(p')- \breve \zeta_m(p)} dp' = \breve R_n(p). \end{equation}
The functions $\breve f_n$ are determined from $\{\breve R_n,\breve \zeta_n,\breve \eta_n\}$ by solving this integral equation. These include the case of a separable sum considered in \cite{NZ20}.

\subsection{Mixed Type Solution}

If we take dressing functions that are linear combinations of the dressing functions produced in the above cases (i.e. we consider $R$ of the form introduced in definition 1), then we can compute nonlinear superpositions of these solutions.
From the above discussions it is easy to write out the appropriate integral equations for the mixed solutions, however they become notationally messy.
We therefore simply mention this possibility, and leave it to the reader to complete the rest.

If the assumptions on $R$ were slightly relaxed to allow a contour $\gamma_m$ in a mixed type solution that is an infinite line, then this class of solutions includes all rapidly decaying solutions to the KdV equation \cite{ZZ19}. 

\subsection{Existence and Uniqueness for Some Cases}

Let $\pi_1(\lambda,\eta) = \lambda$ and $\pi_2(\lambda,\eta) = \eta$.
The {\it nonlocal overlap} condition is the condition that
\begin{equation} \pi_1 (\text{supp}(R(\lambda,\eta))) \cap \pi_2 (\text{supp}(R(\lambda,\eta))) = \emptyset. \end{equation}
In the real case, if there is no {\it nonlocal overlap} then the kernel is a regular Hilbert kernel and the integral equation therefore has a unique solution.
In singular cases, such as the finite gap solutions to the KdV equation, this argument doesn't work.
However, it is likely possible to use a limiting argument to produce a solution in the singular case.
Then the only issue would be verifying the Fredholm alternative.

When there is a nonlocal overlap and the solution is supported and loops and arcs, then a nonlocal overlap can sometimes be solved using a Riemann--Hilbert problem. For example, this is the case when one produces the finite gap solutions to the KdV equation.

\section{Solutions to the KP and (2+1)D KB Equations by the Dressing Method}

In the case that
\begin{equation} R(\zeta,\eta,x,y,t) = e^{\phi(\eta,x,y,t)-\phi(\zeta,x,y,t)} R_0(\zeta,\eta) \end{equation}
with $\phi(\zeta,x,y,t) = \zeta x + \zeta^2 y + \zeta^3 t$ (there is an error here, see footnote \footnotemark[\value{footnote}] on page 4) then 
\begin{equation} u(x,y,t) = \frac{2}{\pi} \frac{\partial}{\partial x} \iint_{\mathbb{C}} f(\xi,x,y,t) d^2 \xi \end{equation}
is a solution to the KP equation
\begin{equation} (4u_t - 6 u u_x + u_{xxx})_x +  u_{yy} = 0, \end{equation}
and the function $\psi(\lambda,x,y,t) = e^{\phi(\lambda,x,y,t)} \chi(\lambda,x,y,t)$ is known as the wave function for the KP equation.
If we consider this as a complex valued PDE of complex variables, then the two real scaling classes of the KP equation exist as dimensional reductions of this equation with $y = \alpha \upsilon$ with $\upsilon \in \mathbb{R}$ and $\alpha = 1$ or $\alpha=i$.

If we instead consider
\begin{equation} \phi(\lambda,t_1,t_2,\dots) = \sum_{n = 1}^\infty t_n \lambda^n \end{equation}
then 
\begin{equation} u(t_1,t_2,\dots) = \frac{2}{\pi} \frac{\partial}{\partial t_1} \iint_{\mathbb{C}} f(\xi,t_1,t_2,\dots) d^2 \xi \end{equation}
is a solution to the KP hierarchy, and $\psi(\lambda,t_1,t_2,\dots) = e^{\phi(\lambda,t_1,t_2,\dots)} \chi(\lambda,t_1,t_2,\dots)$ is a wave function for the KP hierarchy.
The KP hierarchy can be used to traverse the isospectral manifold of all finite gap or Bargmann potentials with the same spectrum.
These isospectral manifolds are finite dimensional.
The isospectral manifolds of periodic and primitive potentials are infinite dimensional.
The KP1 and KP2 are the two scalings of the real KP equation, and are the two real reductions of the complex KP equation.
The KdV is a dimensional reduction of the KP equation.

In the case that $\phi(\lambda,u,v,t) = \lambda u + a \lambda^{-1} v + (\alpha \lambda^2+\beta \lambda^{-2}) t$ then 
\begin{equation} \varphi(u,v,t) = -\log\left(1- \frac{1}{\pi} \iint_\mathbb{C} \frac{f(\xi,u,v,t)}{\xi} d\xi \right),\end{equation}
\begin{equation} \eta = -\frac{1}{\pi}\frac{\partial}{\partial v} \iint_{\mathbb{C}} f(\xi,u,v,t) d\xi - \frac{1}{2}\varphi_{uv}(u,v,t) \end{equation}
solves the (2+1) dimensional completely integrable generalization of the Kaup--Broer system
\begin{align}
&\varphi_t + \alpha \varphi_u^2-\beta \varphi_{v}^2 + \Pi = 0 \\
& \eta_t+2\alpha (\varphi_u \eta)_u - 2\beta(\varphi_v \eta )_v  - 2 a \alpha \varphi_{uu} + 2 a \beta \varphi_{vv} + \frac{\alpha}{2}  \varphi_{uuuv} - \frac{\beta}{2} \varphi_{uvvv} = 0 \\
 & \Pi_{uv} = 2 \alpha \eta_{uu} - 2 \beta \eta_{vv}.
\end{align}
This one complex equation has dimensional reductions to all four scaling classes of the real Kaup--Broer system in (1+1)D
\begin{align}
&\varphi_t + \frac{1}{2}  (\varphi_x)^2 + \varepsilon \eta = 0, \\
&\eta_t + \varphi_{xx} + (\eta \varphi_x)_x + \mu \varphi_{xxxx} = 0,
\end{align}
were $\varepsilon = \pm 1$ and $\mu = \pm 1/4$.

If we instead consider
\begin{equation} \phi(\lambda,u_1,v_1,u_2,v_2, \dots) = \sum_{n = 1}^\infty u_n \lambda^n + v_n \lambda^{-n}. \end{equation}
then
\begin{equation} \varphi(u_1,v_1,\dots) = -\log\left(1-\frac{1}{\pi} \iint_\mathbb{C} \frac{f(\xi,u_1,v_1,\dots)}{\xi} d\xi \right), \end{equation}
\begin{equation} \eta(u_1,v_1,\dots) = -\frac{a}{\pi} \frac{\partial}{\partial v_1} \iint_{\mathbb{C}} f(\xi,u_1,v_1,\dots) d\xi - \frac{a}{2} \varphi_{u_1 v_1}(u_1,v_1,\dots) \end{equation}
is a solution to the full complex (2+1)D Kaup--Broer hierarchy.
The Kaup--Broer hierarchy allows the manifolds of certain  2D Schr\"{o}dinger equations with the same rational spectral curves to be computed. In this case the isospectral manifolds are manifolds of 2D Schr\"{o}dinger operators with electromagnetic fields at fixed level.

\section{Conclusions}

In this note we introduced a family of simple singular linear integral equations that can be used to solve many cases of the nonlocal $\overline{\partial}$-problem.
These solutions allow the calculations of large families of solutions to the KP equation and a (2+1)D generalization of the KB system, which we describe explicitly. In principle, the method discussed in this note will be applicable to any (2+1)D completely integrable equations for which the dressing function is known.

It has already been shown how finite gap solutions to the KdV equation can be produced using the approach to the dressing method discussed in this note. A natural avenue of future work is to figure out how to produce finite gap solutions to the KP equation -- and other (2+1) dimensional completely integrable systems -- using this approach.

\appendix

\section{Classical Examples of Local Solutions to Linear PDEs}

Classically, mathematicians studied ordinary and partial differential equations from the point of view of calculating local solutions. That is, they were looking for formulas for analytic functions that solve the equations. Due to the advent of computers, it has become popular to think of ODEs and PDEs in terms of functions spaces of solutions. One of the most elementary classical solution is the local solution is d'Alembert's solution to the (1+1)D wave equation.

The (1+1)D wave equation {$u_{tt} = c u_{xx}$} has the local d'Alembert solution
\begin{equation} u(x,t) = f(x-ct)+g(x+ct). \end{equation}
In imaginary velocity case $c = i$, this equation becomes the elliptic 2D Laplace equation $u(x,y) = f(x+iy) + g(x-iy)$ where $f(w)$ and $g(w)$ solve the Cauchy--Riemann equations
\begin{equation} \frac{\partial f}{\partial \overline{w}} = 0. \end{equation}

This approach generalizes to Whittaker's solutions to the (2+1)D wave equation $u_{tt} = c^2 \Delta u$ and the 3D Laplace equation $\Delta u = 0$ is the case of imaginary time \cite{W03,WW96}.
Whittaker's solution to the (2+1)D wave equation is
\begin{equation} u(x,y,t) = \int_0^{2\pi} f(\cos(\theta)x + \sin(\theta)y + ct, \theta) d \theta \end{equation}
and Whittaker's solution to the 3D Laplace equation is
\begin{equation} u(x,y,z) = \int_0^{2\pi} f(\cos(\theta)x + \sin(\theta) y + i z, \theta) d \theta \end{equation}
where $f(w,\theta)$ solves the Cauchy--Riemann equations
\begin{equation} \frac{\partial f}{\partial \overline{w}} (w,\theta) = 0. \end{equation}
Expanding $f$ as a Tylor series in $w$ and a Fourier series in $\theta$ gives
\begin{equation}  f(z,\theta) = \sum_{n,m=-\infty}^\infty a_{nm} w^m e^{i n \theta}, \end{equation}
and therefore the general solution to the Laplace equation is the Whittaker series
\begin{equation} u(x,y,z) = \sum_{n,m=-\infty}^\infty {a_{nm}} \int_{0}^{2\pi} \left(\cos(\theta) x + \sin(\theta) y + i z \right)^m e^{in\theta} d \theta. \end{equation}
The Whittaker series solution to the (2+1)D wave equation is
\begin{equation} u(x,y,t) = \sum_{n,m=-\infty}^\infty {a_{nm}} \int_{0}^{2\pi} \left(\cos(\theta) x + \sin(\theta) y + ct \right)^m e^{in \theta} d \theta. \end{equation}
This series solution is a linear function of the coefficients $a_{nm}$.
The Whittaker series allows the computation of solutions without boundary conditions.

The Whittaker series also generalize to the wave equation and Laplace equation in higher dimensions.
The Whittaker series has the down side that the coefficients $a_{nm}$ do not uniquely determine the solution, however it was used by Whittaker to justify the classical series solutions to the Laplace equation (series involving trigonometric functions, Bessel functions, spherical harmonics, etc.) before the advent of spectral theory. The usual spectral theory does not apply to nonlinear equations, so a local solution to the KP equation could help unify the inverse scattering transform for various boundary conditions.

\section*{Acknowledgements}
I would like to thank the reviewers for their insightful suggestions for this paper. The reviewers comments helped improve the paper.

\section*{Conflict of interest}
The author declares that they have no conflict of interest.



\end{document}